\documentclass[aps,pra,twocolumn,showpacs,amsmath,amssymb,amsfonts,superscriptaddress,floatfix,nofootinbib]{revtex4-1}
\usepackage{epsfig} \usepackage{graphicx}
\usepackage{color}
\usepackage[pdftex,bookmarks]{hyperref}

\newcommand{\bgar}{\begin{eqnarray}}
\newcommand{\enar}{\end{eqnarray}} 
 
 \newcommand{\be}{\begin{equation}}
\newcommand{\ee}{\end{equation}}  
 \def\mincirc{\lower
  3pt\hbox{$\buildrel<\over{\hbox{$\mathchar"218$}}$}}
\newcommand{\eotvos}{E$\ddot{\rm o}$tv$\ddot{\rm o}$s}

\begin{document}

\title{\bf Fundamental limitations to  {high-precision} tests of the universality of free fall by dropping atoms\\}

 \date{January 22, 2016}

\author{Anna M. Nobili}
\affiliation{
Dept. of Physics ``E. Fermi'', University of Pisa, Largo B. Pontecorvo 3,
56127 Pisa, Italy}
\affiliation{INFN-Istituto Nazionale di Fisica Nucleare, Sezione di Pisa, Largo B. Pontecorvo 3,  56127 Pisa, Italy}

\begin{abstract}

Tests of the universality of free fall and the weak equivalence principle probe the foundations of General Relativity. Evidence of a violation may lead to the discovery of a new force. The best torsion balance experiments have ruled it out to $10^{-13}$. Cold-atom drop tests have reached $10^{-7}$ and promise to do $7$ to $10$ orders of magnitude better, on the ground or in space. 
They are limited  by the random shot noise, which depends on the number $N$ of atoms in the clouds  (as  $1/\sqrt{N}$).  As mass-dropping experiments in the non-uniform gravitational field of Earth, they are sensitive to the initial conditions. 
 {
Random accelerations due to initial condition errors of the clouds  are designed to be at the same level as shot noise, so that  they can be   reduced with the number of drops along with it. This  sets  the requirements for the initial position and velocity spreads of the clouds with given $N$.  In the STE-QUEST space mission proposal  aiming at $2\cdot10^{-15}$ they  must be about a factor $8$ above the limit established by  Heisenberg's uncertainty principle, and the integration time required to reduce both errors is $3$ years, with a mission duration of $5$ years.}
 { Instead,} offset errors  at release between position and velocity of different atom clouds are systematic and give rise to a systematic effect which mimics a violation. Such systematic offsets must be demonstrated to be as small as required in all drops, i.e., they must be kept small by design, and they must be measured. 
  {For STE-QUEST to meet its goal they  must be several orders of magnitude smaller than the size --in position and velocity space-- of each individual cloud, which in its turn must be at most $8$ times larger than the  uncertainty principle limit.}
  {Even if all technical problems are solved and different atom clouds are released with negligible systematic errors, still these errors must  be measured; and Heisenberg's  principle dictates that such measurement   lasts as long as the experiment.}
While  shot noise is random,  hence its reduction becomes  apparent as more and more drops are performed,  the systematic effect due to offset errors at release  must be identified  through its specific known signature, and measured in order  to be distinguished with certainty from the signal. This requires many well designed measurements to be performed --each to the target precision--  for it to be ruled out  as  a source of violation.
Ways may be pursued in order to mitigate the limitations identified here.

\end{abstract}

\pacs{37.25.+k 04.80.Cc}

\maketitle

\section{Introduction}
\label{sec:Intro}

General Relativity (GR)  rests on the experimental fact that in a gravitational field all bodies fall with the same acceleration regardless of their mass and composition. This is known as the ``Universality of Free Fall'' (UFF), and it is also referred to as the ``Weak Equivalence Principle'' (WEP) though  it is by no means a \textit{Principle} of physics but rather a fact of nature that all experiments, from Galileo  till the present time, have confirmed\cite{NobiliAJP2013}. Experimental evidence of a violation would result in  a scientific revolution, because in such a case either GR must be amended or a new force of nature (\textit{fifth force}) is at play.

UFF/WEP  in the field of Earth  has been tested with macroscopic proof masses of different composition  by dropping them from a height and by suspending them on a torsion balance. The dimensionless parameter  $\eta=\Delta a/a$  which quantifies a violation is obtained by measuring the differential acceleration $\Delta a$ of the proof masses relative to each other  ($\Delta a $ is  the physical observable quantity) as they fall with an average acceleration $a$ towards the Earth ($a$ is referred to as  \textit{driving acceleration}). If UFF/WEP holds,  $\eta=0$; the smaller the value of $\eta$, the more sensitive is the experimental test. In the field of Earth the driving acceleration is $\simeq9.8\,\rm ms^{-2}$ for mass dropping tests and  $\simeq1.69\cdot10^{-2}\,\rm ms^{-2}$ at most (at $45^{\circ}$ latitude) for proof masses suspended on a torsion balance. For the same  sensitivity to differential accelerations, mass dropping tests would yield a smaller value of $\eta$, i.e. a better UFF/WEP test,  by almost a factor $600$.

In spite of this big advantage,  drop tests have measured  $\eta\simeq7\cdot10^{-10}$~\cite{ErseoGAL} while slowly rotating torsion balances~\cite{Adel2008}  have  done $4$ orders of magnitude better, reaching $\eta\simeq10^{-13}$, and finding no violation.  These figures show a higher sensitivity to differential accelerations of the torsion balance {as compared  to mass dropping apparata} by about $4$ million times.  The parameter $\eta$ is named ``\eotvos\ parameter'' in honour of the Hungarian physicist  Roland von \eotvos,  who first used the torsion balance for testing the universality of free fall.

If UFF holds the proof masses fall with the same acceleration. However, if their centers of mass at initial time happen to be located at different heights relative to the center of mass of the source body, a classical differential acceleration arises due to the fact that the gravitational field is not uniform (there is a non zero gravity gradient, or tidal differential acceleration), which mimics a violation. Drop test~\cite{ErseoGAL} is unique (and the best) among drop tests in that the  proof masses are  coupled to one another as two halves (made of $\rm Al$ and $\rm Cu$ respectively)   of a single vertical disk, as suggested  by E. Polacco. During the fall the disk  is sensitive only  to differential accelerations between the centers of mass of its two halves. Should the Earth attract them  differently (\textit{violation}), the disk would rotate about the horizontal axis, an effect that  the authors could measure  very precisely by means of a modified Michelson  {laser} interferometer  in which the two arms terminate on two corner-cube reflectors mounted on the rim of the disk. Ideally, this  is a null experiment: no differential effect $\Rightarrow$ no signal. However, it still depends upon release errors (the disk cannot be dropped with exactly zero rotation rate), which turn out to be the limiting factor, as the authors themselves show~\cite{ErseoGAL}. 

 In the torsion balance the proof  masses are coupled and there is a position of equilibrium which depends on the  design and construction properties of the balance and its suspension fiber, not on the  initial conditions. At $1$-$g$ only  forces  that act on the proof masses along directions which are not parallel to each other  (as it would happen in case of  violation) do affect the balance by causing a non zero torque along the fiber  --which is aligned with the local gravitational acceleration $\vec g$--  thus  displacing the  equilibrium position. The Earth's  gravity gradient  gives a spurious torque along the fiber only by coupling to imperfections in the geometry of the balance, which results in a difference in the directions of its effect  on the proof masses~\cite{TorsionBalanceFocusIssue}.
 
At  a first glance totally free proof masses appear  as the best choice to test  if  they fall with exactly the same acceleration in the field of Earth, both on the ground and in space. Since the proof masses must be sensitive to extremely small  differential forces, the weaker their coupling the better, and one may be tempted to push weak coupling to the limit of no coupling at all (i.e., totally free masses). For macroscopic masses this issue has been thoroughly investigated by many authors~\cite{BlaserReleaseErrors,Tides2003,IorioICE,EPwithlaserranging,NobiliBern2013}, including J. P. Blaser~\cite{BlaserReleaseErrors}, who was the leading scientist of the STEP (Satellite Test of the Equivalence Principle) studies when this proposal was selected by ESA (European Space Agnecy)   as a  medium  size  candidate mission for two times, first in collaboration with NASA and then as an ESA only mission~\cite{STEP}.   All studies~\cite{BlaserReleaseErrors,Tides2003,IorioICE,EPwithlaserranging,NobiliBern2013} have  demonstrated beyond question  that drop tests are not at all the best choice,  on the ground as well as in space, because the initial condition errors  in combination with gravity gradient result in a large systematic effect which severely limits these experiments. 

{As shown in~\cite{EPwithlaserranging},  a nice favourable case is that  of the Moon and the Earth freely falling in the field of the Sun (with an average acceleration $g_{\odot}\simeq6\cdot10^{-3}\,\rm ms^{-2}$),   the Moon's orbit being measured by laser ranging to corner cube reflectors left by astronauts on its surface. In this case the gravity gradient from the Sun is very small thanks  to its very large distance ($d_{\oplus\odot}\simeq1.5\cdot10^{11}\,\rm m$)  yielding $\gamma_{\odot}\simeq\frac{2g_{\odot}}{d_{\oplus\odot}}\simeq1.3\cdot10^{-11}g_{\odot}/\rm m$. Thus, laser ranging with centimeter precision has been able to reach $10^{-13}$(\cite{LLR2012,Mueller2012})  (compatible with $\frac{\gamma_{\odot}}{g_{\odot}}\cdot10^{-2}\simeq1.3\cdot10^{-13}$).  One order of magnitude improvement is expected with the APOLLO laser ranging system at millimiter level~\cite{Apollo2013} once the physical model has been improved accordingly. Beyond that, it will be extremely hard to overcome the effect of gravity gradient and initial condition errors.
}

In recent years, several  tests of UFF/WEP have been performed by dropping atoms in light pulse atom interferometers~\cite{coldatoms2004,OneraWEP2013,RaselEP2014}.
They have reached  $\eta\simeq10^{-7}$, a factor $140$ worse than drop test~\cite{ErseoGAL} and $6$ orders of magnitude worse than the torsion balance test~\cite{Adel2008}, but scientists promise to do many orders of magnitude better,
 on the ground~\cite{Kasevich2007}
or in space~\cite{STEQUEST,QWEPscienceDoc,TinoQWEP2013,STEQUEST2014}. An additional cold atom test, also to $\eta\simeq10^{-7}$, has been published in 2014~\cite{TinoEP2014} but it  is not considered in this work because it is not based on a mass dropping approach, hence initial condition errors as discussed here do not apply to it.

The effect of initial condition errors as studied so far~\cite{BlaserReleaseErrors,Tides2003,IorioICE,EPwithlaserranging,NobiliBern2013} was concerned   only with  macroscopic proof masses. In this work  we revisit the issue  to include atoms dropped  in  light pulse atom interferometers, motivated by the fact that the number of atoms at detection  is very small  compared to Avogadro's number, and therefore the  extremely small mass of the falling bodies plays a key role when considering limitations on position and velocity errors  imposed by  Heisengerg's position-momentum uncertainty principle.

The paper is organized as follows.

 Sec.\,\ref{sec:GradientICE} recalls the basic mathematical formulae for the effect of initial condition errors  in the measurement of the Earth's  gravitational acceleration with atom interferometers. Sec.\,\ref{sec:RandomSystematicICE} analyzes the random and systematic effects of initial condition errors  in cold-atom drop tests of the universality of free fall. Sec.\,\ref{sec:ICEandHP} shows how Heisenberg's  principle limits such tests (due to the  small number of atoms if compared to Avogadro's number) and points out the difference in dealing with  initial condition errors versus shot noise, because the latter is random while the former give rise  also to a systematic differential acceleration  which mimics the signal and must therefore be distinguished from it. 
In the last section we briefly summarize  the results and conclude that  at present cold-atom drop tests of UFF/WEP are not competitive  with results achieved on the ground by torsion balances and with goals pursued in space, all using macroscopic proof masses and  none being based on a mass dropping approach. We also offer some indications as to how the limitations due to initial condition errors outlined in this work can be mitigated (or avoided) in order to improve the present level of UFF/WEP tests  with cold atoms.

 \section{Effects of Earth's gradient and initial condition errors}
\label{sec:GradientICE}

In 1995, by monitoring the motion of a freely falling corner-cube retroreflector  with a  laser interferometer  scientists were able to measure the absolute value of the  local gravitational acceleration  $g$  to $\Delta g/g\simeq1.1\cdot10^{-9}$~\cite{FG51995}. 

In 1999 the absolute value of $g$ was measured to $\Delta g/g\simeq3\cdot10^{-9}$ by dropping caesium atoms in a light pulse  atom interferometer~\cite{PetersNature1999,PetersMetrologia2001}. In this case the key optical elements of the interferometer (beam splitters and mirrors) are implemented by using stimulated Raman transitions between atomic hyperŽfine groundstates (see \cite{PetersThesis1998} for details). The atomic wave packet is split, redirected and finally recombined via atom-light interactions.  The phase that the atoms acquire during the interferometer sequence is proportional to the gravitational acceleration $g$  that they are subjected to. At present shot noise is  the limiting factor, and it is proportional to  $1/\sqrt{N}$, with $N$ the number of atoms at detection. Being random, shot noise is expected to decrease  as  $1/\sqrt{n}$, with $n$ the number of drops.  In a well designed experiment any other noise source that needs to be reduced as $1/\sqrt{n}$ should not exceed the  shot noise limit. If so, the number of drops required to reduce shot noise to the target precision will bring all other noise sources  below the target  too.

The gravity gradient of Earth, combined with  the initial position and velocity of the falling atoms, gives rise to a systematic spurious  effect on their acceleration  (hence on the measured phase shift) which   cannot be neglected if one aims at measuring $g$ to about $10^{-9}$.  For atoms  falling in an atom interferometer this effect has been   calculated by~\cite{PetersThesis1998,WolfTourrenc1999,PetersMetrologia2001} following   the tutorial~\cite{StoreyCohenTannoudji1994}.
Only the contribution  to first order in the gravity gradient $\gamma$ is relevant  and in~\cite{PetersNature1999} it has been  reported to be:
\begin{equation}\label{eq:PetersInitialConditionErrors}
 \Delta g=\gamma \left(\frac{7}{12}gT^{2} -v_\circ T - z_\circ\right) 
\end{equation}
where $T$ is the time interval between successive  light pulses (up to  $ 160\,\rm ms $ in this experiment), $ \gamma\simeq3\cdot10^{-7}\, g/\rm m $ is the gravity gradient in  the laboratory, $ z_\circ$ and $v_\circ $ are the initial position and velocity of the atom. Note that because of a misprint, in~\cite{PetersNature1999} the second term  on the right hand side of  (\ref{eq:PetersInitialConditionErrors}) reads   $ -\gamma v_\circ  $, while it should be multiplied by  $T$.

For a free falling point mass (including one single atom) whose  initial conditions  --nominally zero--  have errors  $\Delta z_{\circ},\Delta v_{\circ}$  (in the direction to the center of mass of Earth)  the first order tidal  acceleration at  the height of fall $z(t)$ is:
\begin{equation}\label{eq:CorrectGradientEffect}
\Delta g(t)=-2\frac{GM_\oplus}{R_\oplus^3}z(t)=\gamma\left(\frac{1}{2}gt^{2} -\Delta v_{\circ}t -\Delta z_{\circ}\right)
\end{equation}
with  $M_{\oplus},\, R_{\oplus}$   the mass and radius of Earth, $G$ the universal constant of gravity and:
\begin{equation}\label{eq:GradientGround}
\gamma=g\frac{2}{R_\oplus}\simeq3.14\cdot10^{-7} g/\rm m
\end{equation}
the  gravity gradient of Earth whose numerical  value is as given by~\cite{PetersNature1999} and reported in (\ref{eq:PetersInitialConditionErrors}).  

It is interesting to note that the disturbing acceleration   (\ref{eq:PetersInitialConditionErrors})  computed  for atoms falling  in the atom interferometer differs from  (\ref{eq:CorrectGradientEffect}) in  the coefficient of the quadratic term   { by $\frac{1}{12}\gamma gT^{2}$. As pointed out by~\cite{Ashby2015}, this discrepancy  is due to the fact that in the atom interferometer the acceleration  of the atoms is  measured as a second difference of their positions at times $0$, $T$ and $2T$ when --during their ballistic flight-- they are subjected to light pulses (see Sec.\,2.1.3 of Peters' PhD thesis\cite{PetersThesis1998}). If we take  (\ref{eq:CorrectGradientEffect}) and integrate twice in order to get the position,  we obtain three position terms proportional to  $t^{4}$,  $t^{3}$ and  $t^{2}$ respectively.  We  then compute the acceleration   {by defining} it as the second position difference at times  $0$, $T$ and $2T$. We find that for the $t^{2}$ and $t^{3}$ terms   the second position difference  is the same as the corresponding  second time derivative, while  this is not so for the $t^{4}$ position term. In this case it is easy to see that the  second  position difference yields  $\frac{7}{12}\gamma gT^{2}$ instead of { $\frac{1}{2}\gamma gT^{2}$},    {with an  acceleration  difference  by $\frac{1}{12}\gamma gT^{2}$~\cite{AEE}. 
 It is apparent that this difference   has nothing to do}  with the ``\textit{quantum} mechanics''  versus ``\textit{classical} mechanics''  approach.  As stated in~\cite{PetersThesis1998} (Sec.\, 2.1.3): \textit{... this type of measurement is not intrinsically ``quantum mechanical''.}  \textit{...We can simply ignore the quantum nature of the atom and model it as a classical point particle that carries an internal clock and can measure the local phase of the light field.}

The acceleration {difference} discussed above does not affect cold-atom drop tests of the universality of free fall, because by taking the  difference of the free fall accelerations of two different atom species or isotopes  --which is the physical quantity to be measured  when testing UFF--  the term quadratic with time in the acceleration    (\ref{eq:PetersInitialConditionErrors}) cancels out. It does not cancel out in experiments to measure the absolute value of $g$, such as~\cite{PetersNature1999}, {in which {case}  the systematic effect (\ref{eq:PetersInitialConditionErrors}) due to initial condition errors in combination with  the Earth's gradient  had to be considerably reduced in order to measure the absolute value of $g$ to $3\cdot10^{-9}$.} The authors report  a careful systematic error analysis  that required to perform many  measurements of $g$ at different heights. By fitting the measurement data to the predicted curve the gradient error could be identified, measured and, to that extent, subtracted (as shown by the authors in their table of systematic effects).

Cold-atom drop tests of the universality of free fall  have been proposed and investigated by the European Space Agency~\cite{QWEPscienceDoc,TinoQWEP2013,STEQUEST,STEQUEST2014} to be performed in space at low Earth altitude.  In these proposals  two overlapped clouds of different isotopes  fall in a Dual-Isotope-Interferometer (DII). The free fall acceleration is measured simultaneously for each cloud. By computing their difference, the acceleration of interest $\Delta g=\eta g(h)$ is derived. 

In absence of weight the leading term measured for each free falling atom cloud is the inertial acceleration arising because of non-gravitational forces  acting on the outer surface of the spacecraft. This inertial acceleration is huge compared to the target, but common to both clouds, and therefore --if the instrument is properly designed-- it can be rejected. If not,  it must be compensated by drag-free control of the spacecraft, which requires a proof mass  unaffected by non gravitational forces (to the level of drag compensation), a sensor to measure its motion relative to the spacecraft, and thrusters (with the necessary amount of propellant) to make the spacecraft follow the proof mass. With  $^{85}{\rm Rb}$,$^{87}\rm Rb$ a  rejection factor of  $4\cdot10^{8}$ is postulated~\cite{STEQUEST2014}. At present the best measured rejection factors are  $550$  for $^{85}{\rm Rb}$,$^{87}\rm Rb$ (Ref.~\cite{OneraWEP2013})  and   $303$ for $^{87}\rm Rb$  and $^{39}\rm K$ (Ref.~\cite{BouyerAirplaneRejection}), showing that an improvement by about $6$ orders of magnitude is needed to meet the requirement.

We mention  for completeness  that on the ground in addition to the gradient (\ref{eq:GradientGround}) there is also  a gradient of the centrifugal acceleration. It is due  to the Earth's daily rotation at angular velocity  $\omega_{\oplus}$, and it adds a  factor  $\leq\omega_{\oplus}^{2}\sim5.4\cdot10^{-10}\,g/\rm m$, which in the $g$ measurement~\cite{PetersNature1999,PetersMetrologia2001}  is negligible.  

At low Earth altitude $h$ the gravity gradient is:
\begin{equation}\label{eq:GammaSpace}
\gamma_{space}=\frac{2}{(R_{\oplus}+h)}\ g(h)/\rm m
\end{equation}
$g(h)$ being the Earth's gravitational acceleration at altitude $h$. Unless the spacecraft attitude is fixed in inertial space the centrifugal gradient must be added {too}, which is $1/2$ of the gravity gradient (\ref{eq:GammaSpace}).

\section{Random and systematic initial condition errors in cold-atom drop tests of UFF}
\label{sec:RandomSystematicICE}

From now on we consider the effect of Initial Condition Errors (ICE) in cold-atom drop tests of UFF/WEP, and  neglect the term quadratic in time because  in the differential acceleration of  two clouds dropped simultaneously  it cancels out. 

Let us start with  one single atom freely falling in the presence of the Earth's gradient $\gamma$. If it has been released at  time $t=0$  with ICE   $\Delta z_{\circ},\Delta v_{\circ}$ (in modulus) the resulting  error at time $t$  in its measured free fall acceleration is:
\begin{equation}\label{eq:DeltagICEsingleatom}
{\Delta g(t)}_{_{ICE-singleatom}}=\gamma\left(\Delta z_{\circ}+\Delta v_{\circ}t\right) 
\end{equation}
where  $\gamma$ is (\ref{eq:GradientGround}) on the ground and (\ref{eq:GammaSpace})   in space. 

If $N$ atoms with these ICE are  released together, random velocities abate with  $\sqrt{N}$, and random position errors are $\sqrt{N}$ smaller too, hence the error in the  acceleration measured at time $t$  is:
\begin{equation}\label{eq:DeltagICEcloud}
{\Delta g(t)}_{_{ICE-singlecloud}}=\gamma\left(\frac{\Delta z_{\circ}}{\sqrt{N}}+\frac{\Delta v_{\circ}}{\sqrt{N}}t\right)\ \ .
\end{equation}
In  the case of a DII, in which the free fall accelerations of two atom clouds are measured independently, each one with a random error  (\ref{eq:DeltagICEcloud}), the random error in their acceleration relative to each other (\textit{differential}) is a factor $\sqrt{2}$ times larger.

Due to the random nature of  noise (\ref{eq:DeltagICEcloud}), it can be reduced by performing many drops, as long as they are uncorrelated. With $n$ such drops the sigmas of the center of mass position and velocity at initial time, i.e.  $\Delta z_{\circ}/\sqrt{N}$ and $\Delta v_{\circ}/\sqrt{N}$ will further decrease as $1/\sqrt{n}$. 

In a DII there is also a  contribution to the differential acceleration  due to position and velocity offset errors $\Delta z_{\circ-rel},\Delta v_{\circ-rel}$  (in the direction to the center of mass of the Earth) between the center of mass position and velocity of the  two clouds at release  {relative to each other}. They arise because of inevitable differences in trapping and  releasing different isotopes/species  --and are therefore systematic-- yielding a systematic differential acceleration:
 \begin{equation} \label{eq:DeltagOffsets}
{\Delta g(t)}_{_{ICE-offsets}}=\gamma\bigg(\Delta z_{\circ-rel} +\Delta v_{\circ-rel}t\bigg)
 \end{equation} 
which mimics a violation.  

Note that an error in the component of the  relative velocity along the orbit does also result in a relative position error in the radial direction (i.e. the direction of a violation signal in the field of the Earth) because different  along track velocities give rise to different orbital radii, the orbital velocity being inversely proportional to the square root of the orbital radius.

It is  mandatory to demonstrate that the measured  signal is not due to the systematic error (\ref{eq:DeltagOffsets}).

A known solution adopted in drop tests with macroscopic proof masses consists in dropping masses of the same composition using  the same apparatus and  performing an experiment as similar as possible to the real one: since in this case there must be no violation, the sensitivity measured sets the level of the UFF test that can be claimed  with this experiment. This check has been done very rigorously in the case of drop test~\cite{ErseoGAL}, and it led to establishing that UFF could not be tested better than  $\simeq7\cdot10^{-10}$ because errors in releasing the vertical disk  could not be reduced below this level.
A null test of this type cannot be done with free identical  atoms as test masses, because identical atoms cannot be distinguished. As suggested by~\cite{MikeHohensee2015}, one should make the two atom clouds slightly different (e.g. by dropping the same atom in different  metastable states), with a difference that allows them to be distinguished in the atom interferometry measurement, but that is negligible for  the sought for UFF violation signal.

An effective alternative when the signature of a systematic effect is known as in this case, is to perform a number of measurements --each to the target precision-- in appropriately modified experimental conditions such that the systematic effect can be separated out  based on its known dependence on the physical parameters involved. In so doing the systematic effect is separated out and measured, so that its contribution to the signal of interest can be firmly identified and possibly reduced below the target. Such a careful analysis of systematics (e.g. as reported by~\cite{PetersNature1999} for the absolute measurement of $g$) obviously requires  the integration time needed to complete one single measurement (by reducing random errors below the target) to be short enough so that systematics can be  separated  from the signal in a realistic time span. This is the case of the torsion balance experiments.

So far cold-atom drop tests of the universality of free fall have been performed on the ground~\cite{coldatoms2004,OneraWEP2013,RaselEP2014} reaching $\eta\simeq10^{-7}$, a factor $140$ worse than drop test~\cite{ErseoGAL} and $6$ orders of magnitude worse than the torsion balance test~\cite{Adel2008}. With  $\gamma\simeq3.14\cdot10^{-7}\,g/\rm m $,  the effect of initial condition errors is not a limitation at this level. 

A cold-atom  drop test  has been proposed in 2007\,\cite{Kasevich2007} aiming   initially at  $\eta=10^{-15}$  and eventually  at  $10^{-17}$,  to be performed      inside  a $10\,\rm m$-tall vacuum chamber built at Stanford University.  They have imaged single clouds~\cite{Kasevich2013}  of  $^{87}\rm Rb$  atoms with  $N=4\cdot10^{6}$  atoms,  $200\,\mu\rm m$ initial radius, $2\,\rm mm/s$  initial velocity spread,   $T=1.15\,\rm s$ and a reported  shot noise limit $\Delta g_{sn}\simeq4\cdot10^{-12}\,g$.  With these values the contribution from ICE to the acceleration error  (\ref{eq:DeltagICEsingleatom}) is reported to produce a phase shift of\ $0.18\,\rm rad$ (Table 1, term 5 in~\cite{Kasevich2013}) which, if compared to the phase shift of\ $2.1\cdot10^{8}\,\rm rad$ produced by the leading $g$ term, yields $\Delta g\simeq8\cdot10^{-10}g$. 
The authors are aware that the  measurement is limited by seismic noise. Nevertheless,  by  comparing various portions of the imaged cloud  and extracting correlated phase noise over many runs they succeed in reducing the phase shift noise  by $\simeq100$, thus inferring an acceleration sensitivity  $\Delta g\simeq6.7\cdot10^{-12}\, g$, close to the shot noise limit. 

The  {``Space Time Explorer and Quantum Equivalence Principle Space Test''} (STE-QUEST) proposal studied by ESA as candidate to a medium size mission~\cite{STEQUEST,TinoQWEP2013,STEQUEST2014} 
aims at a UFF  test to $\eta=2\cdot10^{-15}$  by dropping atoms in a dual isotope $^{85}\rm Rb$,$^{87}\rm Rb$  interferometer in low Earth orbit at $h\simeq700\,\rm km$ altitude {(where $g(h)\simeq8\,\rm ms^{-2}$ and $\gamma(h)\simeq2.83\cdot10^{-7}\,g(h)/m$). 
{It is expected to be limited by a random shot noise differential acceleration  $\Delta g_{sn}\simeq3.66\cdot10^{-13}\,g(h)$  defined as (see~\cite{STEQUEST2014} p.11):
\begin{equation} \label{eq:snSTEQUEST}
\begin{split}
\Delta g_{sn}=\sqrt{2}\frac{1}{C}\frac{1}{kT^{2}}\frac{1}{\sqrt{N}}\\\simeq2.93\cdot10^{-12}\,{\rm ms^{-2}}\simeq3.66\cdot10^{-13}\,g(h)
\end{split}
 \end{equation} 
 where $\lambda=780\,\rm nm$ is the laser wavelength, $ k=\frac{8\pi}{\lambda}$ exploits the technique for enhancing the area of the interferometer,  $T=5\,\rm s$ is the free evolution time (time interval between subsequent light pulses and  $C=0.6$ is the contrast.
 }

{
By performing $n=1.48\cdot10^{5}$ drops, uncorrelated and in the same experimental conditions, the random shot noise (\ref{eq:snSTEQUEST}) can be reduced to $9.5\cdot10^{-16}\,g(h)$ which is a factor $2.1$ below the target violation signal $2\cdot10^{-15}\,g(h)$. In the experiment design outlined in~\cite{STEQUEST2014} --$20\,\rm s$ repetition time, $0.5\,\rm hr$ out of $16\,\rm hr$  dedicated to the experiment at perigee-- one measurement  with this number of drops requires $3$ years to be completed, within a total mission duration of  $5$ years.   
}

{For the random acceleration error (\ref{eq:DeltagICEcloud}) not to exceed the shot noise limit
(\ref{eq:snSTEQUEST}) the atom clouds are required to have 
}
 $N$=$10^{6}$ atoms at detection with 
\begin{equation}\label{eq:STEQUESTclouds}
 300\,\mu{\rm m}\  {\rm initial\ radius}\  ,\ 82\,\mu{\rm ms^{-1}}\  {\rm initial\ velocity\ spread}
 \end{equation} 

\section{Heisenberg's  principle and integration time;  shot noise vs systematic release errors}
\label{sec:ICEandHP}

Each test mass, as well as a single atom,  must obey Heisenberg's uncertainty  Principle (HP), which states:
   \begin{equation}\label{eq:HeinsenbergPrincipleSingleAtom}
  \Delta p_{\circ}\cdot\Delta z_{\circ}\geq\frac{\hbar}{2}
 \end{equation} 
 where  $\hbar=1.054\cdot10^{-34}\,\rm Js$ is the reduced Planck constant and the linear momentum $\Delta p_{\circ}$   contains the mass of the body. For a single atom, because of its extremely small mass, HP requires  (in the case of  $\rm Rb$):
   \begin{equation}
   \begin{split}
\left(\Delta v_{\circ}\cdot\Delta z_{\circ}\right)_{HP-atom}\geq\frac{\hbar}{2} \frac{1}{m_{Rb}}\\ 
\geq\frac{\hbar}{2}  \frac{1}{85.468\cdot10^{-3}} \cdot{N_A}
\ \rm m^{2}/s\\
\geq3.7\cdot10^{-10} \ \rm m^{2}/s  
   \end{split}
\label{eq:AvogadrosRevenge}
 \end{equation} 
 where $m_{Rb}=\frac{8.5\cdot10^{-2}}{N_{A}}\,\rm kg$ is the mass of a Rb atom and  $N_{A}=6.022\cdot10^{23}$ is Avogadro's number.

For each cloud made of  $N$=$10^{6}$ \  $\rm Rb$ atoms released together, the random errors on the initial center of mass velocity and position are reduced by $\sqrt{N}$.  This is equivalent to a free mass with position error  $(\sqrt{N}\Delta z_{\circ})$ and  momentum error $(m_{Rb}\sqrt{N}\Delta v_{\circ})$, for  which HP requires  $\left(Nm_{Rb}\Delta v_{\circ}\cdot\Delta z_{\circ}\right)_{HP-freemass}\geq\hbar/2$, hence:
 \begin{equation}
 \begin{split}
\left(\Delta v_{\circ}\cdot\Delta z_{\circ}\right)_{HP-freemass}
\geq\frac{\hbar}{2}  \frac{1}{85.468\cdot10^{-3}} \cdot\frac{N_{A}}{N}
\ \rm m^{2}/s\\
\geq3.7\cdot10^{-16} \ \rm m^{2}/s\ .
 \end{split}
\label{eq:AvogadrosRevenge2}
 \end{equation} 
 This is the ultimate limit, since it is the HP limit for a single, free Bose-Einstein-Condensate of $N$ atoms, and as such a lower limit to the initial conditions of the real experiment (free atoms released from an optical trap).

In order to quantitatively assess  the implications of Heisenberg's uncertainty principle in cold-atom drop tests of UFF we refer  to the STE-QUEST proposed space experiment because  it has been  investigated within ESA for several years and literature is available that provides all the information needed for quantitative assessment~\cite{STEQUEST,TinoQWEP2013,STEQUEST2014}. A similar experiment was investigated  by ESA  (with participation by the author) to be performed on the International Space Station~\cite{QWEPscienceDoc}. However, the target was less ambitious than for STE-QUEST,  and  the ESA report of that study  is an unpublished draft. 
STE-QUEST  aims at $\eta=2\cdot10^{-15}$, a target that  would improve the current best tests performed with macroscopic masses by a factor  of $50$, which makes the proposal worth considering despite the huge gap (by a factor  of $50$ million) which separates it from the level that atom interferometers  have achieved so far~\cite{coldatoms2004,OneraWEP2013,RaselEP2014,TinoEP2014}. 

In the {STE-QUEST atom clouds} (\ref{eq:STEQUESTclouds})  each atom obeys the HP limit (\ref{eq:AvogadrosRevenge}) by a factor $66$, that is the product of its position and velocity errors is above the uncertainty limit by a factor $66$. Their centers of mass have position and velocity errors  smaller by a factor $\sqrt{N}=10^{3}$, hence they are above  the HP limit (\ref{eq:AvogadrosRevenge2}) by the same factor $66$.
  With a comparable  share of error in position and velocity it means that each error is roughly a factor $\sqrt{66}\simeq8$ above its HP limited value.
{By comparison, position and velocity errors of the single clouds realized  by~\cite{Kasevich2013}  are a factor $\sqrt{1.1\cdot10^{3}}\simeq33$ above the HP limit.
}

In every drop the contribution   (\ref{eq:DeltagICEcloud}) of random ICE   to the acceleration difference  between the  clouds  amounts, {with the planned  initial condition errors (\ref{eq:STEQUESTclouds}), 
 to $\sqrt{2}\cdot{\Delta g}_{_{ICE-singlecloud}}\simeq2.84\cdot10^{-13}\,g(h)$, which is slightly smaller, by a factor $1.3$, than the expected shot noise limit (\ref{eq:snSTEQUEST}).} 
{Thus,  the same number of drops  that need to be performed in order to reduce the shot noise a factor $2.1$ below the target  signal will reduce (also as $1/\sqrt{n}$) the initial random errors --hence the random differential acceleration error $\sqrt2\cdot{\Delta g}_{_{ICE-singlecloud}}$-- to a value $2.7$ times smaller than the signal.}

{With the same number of atoms, were it possible to run the experiment with  position and velocity errors at exactly the limit of  Heisenberg's uncertainty principle, they would result in a random differential acceleration error a factor $\sqrt{66}\simeq8$ smaller, of $3.5\cdot10^{-14}\,g(h)$, which is the lowest possible  limit   and a factor $10$ below the expected shot noise (\ref{eq:snSTEQUEST}).
}

Let us now consider the systematic error (\ref{eq:DeltagOffsets}).

In the list of STE-QUEST systematic errors the requirements set by the proposers for the offset errors  $\Delta z_{\circ-rel}, \Delta v_{\circ-rel}$ at release are (see~\cite{STEQUEST2014},  Table 4, first entry): 
\begin{equation} \label{eq:Offsets}
\Delta z_{\circ-rel}=1.1\,{\rm nm} \ \ \ \  \Delta v_{\circ-rel}=0.31\,\rm nm/s\ \  .
 \end{equation} 

{First of all, it is interesting to compare these requirements  with the position offsets between the centers of mass  as required in  UFF/WEP  experiments using macroscopic proof masses}.

{It was pointed out by~\cite{BlaserReleaseErrors} that GG~\cite{GGFocusIssue2012} is the only proposed space experiment in which the test masses are coupled and motion occurs around a position of relative equilibrium independent from initial conditions (as in the case of the torsion balance). In GG the proof  masses are coaxial cylinders rotating around the symmetry  axis, weakly coupled in 2D (the plane perpendicular to the spin/symmetry axis) whose physical  property of self-centering (starting from construction/mounting offsets of $10\,\mu\rm m$) makes the gradient effect compatible with a  test 200 times   more sensitive than STE-QUEST, to  $\eta=10^{-17}$.}
{In Microscope~\cite{MicroscopeFocusIssue2012}, to be launched in April 2016,  the coaxial test cylinders are sensitive along the symmetry axis while rotation occurs along an axis perpendicular to it; the position offsets required by construction/mounting amount to  $20\,\mu\rm m$, to be reduced to $0.1\,\mu\rm m$ by offline data analysis (over various measurements) based on the specific, known signature of the gradient effect so as to bring it below a target which is two times more sensitive  than that of STE-QUEST, i.e.\, $\eta=10^{-15}$.}

In STE-QUEST, if the requirements (\ref{eq:Offsets})   are met in every drop (though they don't need to be measured to this level in every drop) the  differential acceleration (\ref{eq:DeltagOffsets}) is a factor $2.7$ below the target signal.

It is crucially important    to verify that the offsets between different atom clouds at release meet the requirements (\ref{eq:Offsets}), and that they do meet them in all drops for the entire duration of the experiment. Assume that a fraction $f<1$ of the required number of drops $n$ has initial offset errors that are larger than required by (\ref{eq:Offsets}), to the extent  that the resulting differential acceleration error (\ref{eq:DeltagOffsets}) is $k>1$ times larger than in the remaining $(1-f)n$ drops which meet (\ref{eq:Offsets}). Then, the resulting average error in the differential acceleration is:
\begin{equation} \label{eq:}
\overline{\Delta g}=\frac{fn\cdot k\Delta g+(1-f)n\cdot\Delta g}{n}=[f(k-1)+1]\Delta g\ .
 \end{equation} 
For instance, if a fraction $f=10\%$ of the drops have offsets that produce acceleration errors $k=21$ times larger than those produced in the remaining $90\%$ of the drops in which   (\ref{eq:Offsets}) are met, the resulting systematic error (\ref{eq:DeltagOffsets}) will be 3 times bigger, which is larger than the target signal and indistinguishable from it with a single measurement.

The size of the atom clouds at initial time (in position and velocity space) is the size of the trapped clouds. It is limited by HP (\ref{eq:AvogadrosRevenge}) (see~\cite{BECbook}), and the centers of mass position and velocity are limited by HP (\ref{eq:AvogadrosRevenge2}).

 When confined, clouds of different isotopes/species have  different size, because of the different physical properties of the atoms, including mass. In STE-QUEST the mass difference alone results in a size difference  of  $3\,\mu\rm m$. In current instruments, estimates of the offset errors at release are many orders of magnitude away from the requirements  (\ref{eq:Offsets}). 
 In the UFF test~\cite{OneraWEP2013}  to $\eta\simeq10^{-7}$  {the estimates reported  are   $\Delta z_{\circ-rel}= \pm 0.2\,\rm mm$ and $\Delta v_{\circ-rel} \leq6\,\rm mm/s$, which are about $1.8\cdot10^{5}$  and $1.7\cdot10^{7}$ times larger, respectively,  than   (\ref{eq:Offsets}).
 } 
 
 Nevertheless, there is no theoretical limit to the accuracy with which the centers of the two clouds can be made coincident. In principle, with enough care in preparing and releasing the trap, the offsets can be made to meet  (\ref{eq:Offsets}). 
 {However, such preparation requires  validation and only repeated measurements in the actually realized trap can provide  it.}
In order to be excluded as the cause of any anomalous acceleration found in the experiment (violation?) the initial offsets  must be measured, and the measurement is limited by HP (\ref{eq:AvogadrosRevenge2}), namely by the uncertainty limit in  position and velocity of the center of mass of each  cloud.

On the other hand, the required initial offsets  (\ref{eq:Offsets})   are well below this limit. Because of the  extremely small mass of the clouds  (even $10^{6}$ atoms are very few  compared to Avogadro's number)  they  are  below the HP limit (\ref{eq:AvogadrosRevenge2}) by a factor  {$\sqrt{1.1\cdot10^{3}}\simeq33$}. Thus, the  uncertainty principle prohibits the initial offsets to be measured to the required precision in a small number of drops. 

Only by measuring them for the entire integration time of the UFF test, and by averaging over as many drops as required for the test,  they can be proven to  meet the requirements. 

Should STE-QUEST aim at testing  UFF only two times better than the present goal, to $\eta=10^{-15}$, it would require an integration time $4$ times longer, of $12$ years, to complete one single measurement  and to measure the initial offsets to the level required! {In addition,  the initial size and velocity spread of each cloud  would have to be only a factor $4$ above the HP limit. These facts explain why the target of STE-QUEST could not possibly be pushed to $10^{-15}$ in order  to make it competitive with Microscope.}

Assuming that all technical problems are solved, and the initial offsets are negligibly small, still they must be measured and the  integration time needed  (for a given target) to reach the precision required is  dictated by   HP limit  (\ref{eq:AvogadrosRevenge2}). This is a fundamental  limit  and can be relaxed only by increasing the number $N$ of atoms in the clouds (without increasing position and velocity errors in their centers of mass), which would as well reduce  the shot noise limit.   

{Should the target of STE-QUEST be one order of magnitude less ambitious, to $2\cdot10^{-14}$, with the same shot noise,  then initial condition errors could  be $10$ times larger,  hence the clouds would be a  much safer factor $\sqrt{665}\simeq82$ above Heisenberg's limit and the integration time would be a factor $100$ shorter, requiring $11$ days. Offset errors at release could  also be $10$ times larger. They would still be below the HP limit for each cloud (but only slightly, by a factor $\sqrt{11}\simeq3.3$), and their measurement would require the same integration time, but this would now be realistic and leave enough time for checking their systematic effect. At this level limitations would not be fundamental but mostly technical, as it is inevitable given the 5 million gap from the current state of the art. However, this goal would be $20$ times less sensitive than Microscope's goal and only $5$ times better than the current best tests of  of UFF/WEP, making  the case for an expensive space mission extremely weak. 
 }

{In its current design} STE-QUEST proposes to measure the offsets at apogee (while drops to test UFF are performed at perigee), by producing atom clouds and imaging them in order to verify, based on their evolution, how far apart they were at release (see~\cite{STEQUEST2014}, p.\,12). For this approach to work, one should demonstrate that systematic errors are the same in both cases, and that the accuracy of the imaging method is close to Heisenberg's limit.

The key difference between random and systematic errors must be kept  clearly in mind. Once a random error has been reduced to a certain level  the result is apparent, and if it has reached the design level the measurement is over. Instead,   a systematic effect which is known to mimic the signal requires a number of different measurements, each of them to the target level, in order to verify its specific signature, i.e., the way it depends on some physical parameters, for  it to be distinguished from the signal beyond doubts. It is well known that a very long integration time rules out the possibility of  a careful check of systematic errors and questions the significance of an experimental result. 

It has been suggested  to cancel  the Earth's gradient by placing a mass nearby. A reduced value of  $\gamma$ would reduce the systematic effect (\ref{eq:DeltagOffsets}) and possibly make it irrelevant, at least when aiming at $\eta=2\cdot10^{-15}$.   Inside the small experimental  region in which atoms are dropped  the Earth's gradient is almost constant, but time varying depending on the orbital motion and the attitude of the spacecraft. Instead, the mass must be fixed  (to avoid bigger problems),  and very close by (at $50\,\rm cm$  more than 2 tons would be required). Hence, its gradient changes across the region but it is constant in time. As a result, the Earth's gradient would  be either under compensated or over compensated, thus not solving the problem. On the ground this difficulty may be reduced because the Earth's gradient in the experimental region does not change with time (the lab doesn't move), and a large mass could be placed far away. However, with the atoms falling at $1$-$g$ the experimental chamber inside which the Earth's gradient must be compensated is certainly larger (as in the $10\,\rm m$ evacuated tower at Stanford).

{Another possibility may be worth investigating. Instead of using position and momentum as conjugate variables (subject to the uncertainty relation
(\ref{eq:HeinsenbergPrincipleSingleAtom})) one may try to combine them in one single variable such that its error can be minimized at the expense of the error in its conjugate. This technique  (known as \textit{squeezing}) has been recently applied to  shot noise in atomic clock measurements~\cite{KasevichSqueezing2016}  with a reduction equivalent to increasing the number of atoms by a factor 100, i.e. equivalent to reducing the phase measurement noise, hence the acceleration shot noise, by a factor $10$.}

\section{Conclusions}
\label{sec:Conclusions}

In experiments to test the universality of free fall and the weak equivalence principle with free macroscopic masses, initial condition  errors are  a well known limitation~\cite{ErseoGAL,BlaserReleaseErrors,EPwithlaserranging}. 
Macroscopic masses have Avogadro's number on their side, while cold-atom drop tests  are limited  by Heisenberg's  principle because of the vanishingly small mass of the atom clouds. 

The proposed  STE-QUEST space experiment~\cite{STEQUEST,TinoQWEP2013,STEQUEST2014} investigated by the European Space Agency needs $3$ years to complete one single measurement of the universality of free fall  to $2\cdot10^{-15}$ within a total mission duration of  $5$ years. It is known that such a very long integration time is set by the need  to reduce the random shot noise. 
 
 Here we have investigated the effects of initial condition errors in the presence of the Earth's gravity gradient, and in particular the systematic differential acceleration due to (systematic)  offset errors at release between clouds of different isotopes/species. We have shown that: 
 
 (i)  the requirements {(\ref{eq:Offsets})} on { position and velocity} offset errors at release  for this systematic differential acceleration not to exceed the shot noise limit are  --for the goal of STE-QUEST-- a factor   { $\sqrt{1.1\cdot10^{3}}\simeq33$} below the limit set by {Heiseberg's} principle;

 (ii)  the systematic offset errors  { (\ref{eq:DeltagOffsets})} must  meet the requirements  in all drops, and this must be ensured by measuring them, which requires --because of Heisenberg's principle limit {for each cloud}-- the same integration time needed to reduce the random shot noise; 
 
(iii) the systematic nature of the effect  caused by offset  errors  at release  demands --for the  experiment to be reliable and its result to be acceptable-- that more measurements are performed until it is possible to distinguish this effect from the target violation signal. 
 
The   integration time set by the uncertainty principle can be reduced only by increasing the number of atoms in the clouds (as long as this is done without increasing their position and  velocity errors), which would as well reduce the shot noise. 
  
The Achille's heel of light pulse atom interferometers  in testing  the universality of free fall  {to $2\cdot10^{-15}$ and better} by dropping atoms  appears to be the extremely small mass of the  atom clouds.

{Were STE-QUEST aiming at a $10$ times less ambitious goal, to $2\cdot10^{-14}$, it would not hit the fundamental limits outlined here. In this case the challenges would be mostly technical, in order to bridge the $5$ million gap which separates this goal from the current $10^{-7}$ level of UFF/WEP   drop tests with cold atoms.}

{When aiming at $2\cdot10^{-15}$ --like STE-QUEST in its current design-- or better,  ways may be pursued  to overcome, or at least to alleviate, the limitations pointed out in this work.} 

Squeezing techniques can be investigated, which would reduce the effect of initial condition errors,  thus allowing the corresponding requirements to be relaxed. 

The very large number of drops needed  seems inevitable; however, one might optimize the  time needed to perform them both in space and on the ground. 

Partial compensation of the Earth's gradient by means of an appropriate artificial mass nearby seems unrealistic in space but may be attempted on the ground with a trade-off between the free fall time (hence the size of the experimental chamber) and a reduced gradient (hence relaxed requirements on initial condition errors).

A zero check by dropping the same atoms with some clever technique to allow them to be distinguished as suggested by~\cite{MikeHohensee2015} can be investigated on the ground. If well designed, no appreciable violation should occur and the experiment would reliably  establish the limiting value of $\eta$  that a cold-atom drop test of UFF can achieve.

Nonetheless, a UFF test to a few $10^{-15}$ by dropping atoms appears to be hard. Cold-atom tests which are not based on mass dropping approach (such as~\cite{TinoEP2014})  should not be affected by initial condition errors and maybe worth
closer attention by the community.
 
As at present, a comparison with space experiments using macroscopic masses and not based on the mass dropping approach shows that Microscope~\cite{MicroscopeFocusIssue2012} (to be launched in April 2016) can make one measurement to $10^{-15}$ in $1.4\,\rm d$ while   GG~\cite{GGFocusIssue2012} requires a few hours to reach $10^{-17}$; the limitation being  thermal noise at room temperature in both cases but  in different frequency regions of the signal due to different up-convertion rates by means of rotation~\cite{PRLthermalnoise,IntegrationTimePRD2014}.   

\vspace{5mm}

\textbf{Acknowledgements.}

The author wishes to thank  G. Catastini, for pointing out the need to check Heisenberg's  principle,  and A. Anselmi  for invaluable discussions. Contributions from    M. Hohensee, E. Adelberger, C. S. Unnikrishnan, N. Ashby,  M. Shao, M. Maggiore, S. Stringari,  E. Polacco, F. Pegoraro and E. Arimondo are gratefully acknowledged. This work is dedicated to the memory of Professor Polacco, who passed away while it was being completed.


\begin{thebibliography}{1}
%
\bibitem{NobiliAJP2013}A. M. Nobili, D. M. Lucchesi, M. T. Crosta, M. Shao, S. G. Turyshev, R. Peron, G. Catastini, A. Anselmi, and G. Zavattini,  On the universality of free fall, the equivalence principle,
and the gravitational redshift, \href{http://dx.doi.org/10.1119/1.4798583} {Am. J. Phys. {81}, 527 (2013)}
\bibitem{ErseoGAL} S. Carusotto, V. Cavasinni, A. Mordacci, F. Perrone, E. Polacco, E. Iacopini and  G. Stefanini, Test of g universality with Galileo type experiment,  {Phys. Rev. Lett. {69}, 1722 (1992)} 
\bibitem{Adel2008} {S. Schlamminger},  K. Y. Choi, T. A. Wagner,  J. H. Gundlach,  and E. G. Adelberger, Test of the equivalence principle using a rotating torsion balance, {Phys. Rev. Lett. 100, 041101 (2008)}
\bibitem{TorsionBalanceFocusIssue} {T. A.  Wagner}, S. Schlamminger, J. H. Gundlach, and E. G. Adelberger, Torsion-balance tests of the weak equivalence principle, Class. Quantum Grav. 29, 184002   (2012)
\bibitem{BlaserReleaseErrors} J. P.  Blaser, Can the equivalence principle be tested with freely orbiting masses?, {Class. Quantum Grav. 18 2509-2514 (2001)}
\bibitem{Tides2003} G. L. Comandi, A. M. Nobili, R. Toncelli, and M. L. Chiofalo, Tidal effects in space experiments to test the equivalence principle:
implications on the experiment design, Phys. Lett. A 318, 251 (2003)
\bibitem{IorioICE} L. Iorio, On the possibility of testing the weak equivalence
principle with artificial Earth satellites, Gen. Relat. Gravit. 36, 361 (2004)
\bibitem{EPwithlaserranging} {A. M. Nobili}, G. L. Comandi, D. Bramanti, S. Doravari,  D. M. Lucchesi and F. Maccarrone,  Limitations to testing the equivalence principle with satellite laser ranging, Gen. Relat. Gravit. 40, 1533 (2008)
\bibitem{NobiliBern2013} {A. M. Nobili}, Testing the weak equivalence principle with macroscopic proof masses on ground and in space: a brief review,   Int. J. Mod. Phys. Conf. Ser. 30, 1460254   (2014) 
\bibitem{STEP} J. P. Blaser {\it  et al.,}   \textit{Satellite Test of the Equivalence Principle}, Report on the Phase A Study, ESA/NASA SCI (93)4 (1993) \& ESA SCI (96)5 (1996)
%
\bibitem{LLR2012} {J. G. Williams, S. G. Turyshev \& D. H. Boggs}, Lunar laser ranging tests of the equivalence principle, Class. Quantum Grav. 29, 184004 (2012)
\bibitem{Mueller2012}{J. Mueller, F. Hoffman \& L. Biskupek}, Testing various facets of the equivalence principle
using lunar laser ranging,  Class. Quantum Grav. 29, 184006 (2012)
\bibitem{Apollo2013} T. W. Murphy, Lunar laser ranging: the millimeter challenge, Rep. Prog. Phys. 76, 076901 (2013) 
%
\bibitem{coldatoms2004} S. Fray, C. A. Diez, T. W. Hansch, and M. Weitz, Atomic interferometer with amplitude gratings of light and its applications
to atom based tests of the equivalence principle, Phys. Rev. Lett. 93, 240404  (2004)
\bibitem{OneraWEP2013}  A. Bonnin, N. Zahzam, Y. Bidel, and A. Bresson, Simultaneous dual-species matter-wave accelerometer, Phys. Rev. A 88, 043615  (2013)
\bibitem{RaselEP2014} D. Schlippert,  J. Hartwig,  H. Albers, L. L. Richardosn,  C. Schubert, A. Roura, W. Schleich, W. Ertmer, and E. M. Rasel, Quantum test of the universality of free fall, Phys. Rev. Lett. 112, 203002 (2014)
%
\bibitem{Kasevich2007} S. Dimopoulos, P.  W. Graham, J. M. Hogan, and M. A. Kasevich, Testing general relativity with atom interferometry, Phys. Rev. Lett.  98, 111102  (2007)
\bibitem{STEQUEST} STE-QUEST space mission proposal:\\ http://sci.esa.int/ste-quest/
\bibitem{QWEPscienceDoc} ESA (European Space Agency)  Contract \\ 4000105593/12/NL/CO, ``Q-WEP:  Atom Interferometry Test of the Weak Equivalence Principle in Space'', Requirements Document, unpublished draft (June 2013) 
\bibitem{TinoQWEP2013} G. M. Tino  {\it  et. al.,} Precision gravity tests with atom interferometry in space, Nuclear Physics B (Proc. Suppl.)  243-244,  203-217 (2013), partly based on Ref.~\cite{QWEPscienceDoc}
\bibitem{STEQUEST2014} D. N. Aguilera  {\it  et. al.,} STE-QUEST--test of the universality of free fall using cold atom interferometry, Class. Quantum Grav. 31,  115010  (2014)
\bibitem{TinoEP2014} M. G. Tarallo, T. Mazzoni, N. Poli, D. V. Sutyrin, X. Zhang, and G. M. Tino,  Test of Einstein equivalence principle for 0-Spin and half-integer-spin stoms: search for spin-gravity coupling effects, Phys. Rev. Lett. 113, 023005 (2014)
\bibitem{FG51995} T. N. Niebauer, G. S. Sasagawa, J. E. Faller, R. Hilt and F. Klopping, A new generation of absolute gravimeters, Metrologia 32, 159 (1995) 
\bibitem{PetersNature1999} {A. Peters,  K. Y. Chung,   and  S. A. Chu}, Measurement of gravitational acceleration by dropping atoms, Nature 400, 849 (1999) 
\bibitem{PetersMetrologia2001} {A. Peters,  K. Y. Chung,  and S. A. Chu}, High-precision gravity measurements using atom interferometry, Metrologia 38, 25 (2001)
\bibitem{PetersThesis1998} A. Peters, {High precision gravity measurements using atom interferometry}, PhD Thesis, Stanford University (1998)
\bibitem{WolfTourrenc1999} P. Wolf \&  P. Tourrenc, Gravimetry using atom interferometers: some systematic effects, Phys. Lett. A  251, 241-246 (1999)
\bibitem{StoreyCohenTannoudji1994} P. Storey \& C. Cohen-Tannoudji, The Feynman path integral approach to atomic interferometry. A tutorial, J. Phys. II France 4, 1999 (1994) 
 \bibitem{Ashby2015} N. Ashby,  \textit{personal communication}  (2015)
 \bibitem{AEE} A. M. Nobili, Comment on  the measurement of the gravitational acceleration by light pulse atom interferometry, \textit{to be submitted} (2016) 
\bibitem{BouyerAirplaneRejection}  G. Varoquaux, R. A. Nyman, R. Geiger, P. Cheinet,  A. Langradin and P. Bouyer, How to estimate the differential acceleration in a two-species atom interferometer to test the
equivalence principle, New J. Phys. 11, 113010  (2009)
\bibitem{MikeHohensee2015} {M. Hohensee}, \textit{personal communication}  (2015)
\bibitem{Kasevich2013} S. M. Dickerson,  J. M. Hogan, A. Sugarbaker, D. M. S. Johnson, and M. A. Kasevich, Multiaxis inertial sensing with long-time point source atom interferometry, Phys. Rev. Lett. 111, 083001   (2013)
\bibitem{GGFocusIssue2012}{A. M. Nobili}, M. Shao, R. Pegna, G. Zavattini, S. G. Turyshev, D. M. Lucchesi, A. De Michele, S. Doravari, G. L. Comandi, T. R. Saravanan, F. Palmonari, G. Catastini, and A. Anselmi, ``Galileo Galilei'' (GG): space test of the weak equivalence principle to $10^{-17}$ and laboratory
demonstrations, Class. Quantum Grav. 29, 184011   (2012)
\bibitem{MicroscopeFocusIssue2012} P. Touboul, G. Metris, V. Lebat, and A. Robert,  The MICROSCOPE experiment, ready for the
in-orbit test of the equivalence principle, Class. Quantum Grav. 29, 184010 (2012)
\bibitem{BECbook} C. J. Pethick and H. Smith, \textit{Bose-Einstein Condensation in Diluted Gases} (Cambridge University Press, 2008)
%
{\bibitem{KasevichSqueezing2016} O. Hosten, N. J. Engelsen, R. Krishnakumar, and M. A.   Kasevich, Measurement noise $100$ times lower than the quantum-projection limit using entangled atoms, Nature doi:10.1038/nature16176 (2016)}
%
\bibitem{PRLthermalnoise} {R. Pegna,} A. M. Nobili, M. Shao, S. G. Turyshev, G. Catastini, A. Anselmi,  R. Spero, S. Doravari, G. L. Comandi, and  A. De Michele, Abatement of thermal noise due to internal damping in
2D oscillators with rapidly rotating test masses, Phys. Rev. Lett. 107, 200801  (2011)
\bibitem{IntegrationTimePRD2014} {A. M. Nobili}, R. Pegna,  M. Shao, S. G. Turyshev, G. Catastini, A. Anselmi,  R. Spero, S. Doravari, G. L. Comandi, and  A. De Michele, Integration time in space experiments to test the equivalence principle, Phys. Rev. D 89, 042005  (2014)

\end{thebibliography}
\end{document}